\documentclass[review]{elsarticle}

\usepackage{hyperref} 
\usepackage{amsmath} 

\journal{Journal of \LaTeX\ Templates}

\begin{document}

\begin{frontmatter}

\title{Innovative Thermo-Solar Air Heater}

\author[mymainaddress,myfourthadress]{M. Cuzminschi}

\author[mymainaddress,mysecondaryaddress]{R. Gherasim}

\author[mymainaddress]{V. Girleanu}

\author[mymainaddress,mythirdadress]{A. Zubarev\corref{mycorrespondingauthor}}
\cortext[mycorrespondingauthor]{Corresponding author}
\ead{alxzubarev@gmail.com}

\author[mymainaddress]{I. Stamatin}

\address[mymainaddress]{University of Bucharest, Faculty of Physics, Atomistilor 405, P.O 38, Bucharest-Magurele, Romania, 077125}
\address[mysecondaryaddress]{Center of Technology and Engineering for Nuclear Projects, 409, Atomistilor Street, Magurele, Judet Ilfov, Romania}
\address[mythirdadress]{National Institute for Laser, Plasma and Radiation Physics, P.O. Box MG-36, Magurele, Bucharest, Romania}
\address[myfourthadress]{Department of Theoretical Physics, IFIN-HH, Magurele, Romania}

\begin{abstract}

In the present work we elaborate the innovative design of the solar air heater and justify it by a Computational Fluid Dynamics (CFD) simulation, implementing and experimentally testing a sample. We propose to use this device for maintenance of constant ambient conditions for thermal comfort and low energy consumption  for indoor environments, inside greenhouses, passive houses, and to protect buildings against temperature fluctuations. We tested the functionality of our sample of the solar air heater for 50 weeks and obtained an agreement between the results of the numerical simulation, implemented using OpenFOAM (an open source numerical CFD software) and the experimental results.
\end{abstract}


\end{frontmatter}


\section{Introduction}

Heating, ventilation and air-conditioning (HVAC) systems are mainly designed for the building sector aiming to ensure the comfort living standards for various climatic zones. The building sector accounts for more than $39\%$ of the primary energy requirements \cite{Building-Heating} and is a main contributor to carbon emission. The Solar-HVAC has been considered as an alternative to reduce contribution of the primary energy and in this respect, are developed  solutions based on the direct transformation of the solar energy in internal energy of the transport medium \cite{SuplimentaryV1,SuplimentaryV2,Solar-Patent1,Air-Heater2}.

This basic principle is successfully applied to the solar air heaters in passive houses keeping a minimal comfortable temperature of $15^oC$ \cite{Indoor-Temperature} during cold seasons.

One niche where the solar-air heaters can bring an input of additional heat during cold season is the greenhouse growers. The cost of fuel is an increasingly significant production expense for greenhouse growers in temperate climates. High heating costs motivate growers to improve the efficiency of crop production to minimize energy inputs. The two parameters influencing plant development are: mean daily/night temperature (MDNT) and photosynthetic daily light integral (DLI)\cite{SuplimentaryV3}.

MDNT usually must stay in a close range of $15-20^oC$ for optimal conditions in the greenhouse environment. During warm season the growers are faced with high energy consumption for seeds and fruits drying, the intermediate stage in preservation. Several experimental demonstrations and reviews show that the electricity from primary energy sources can be replaced with solar energy (thermal and photovoltaic \cite{BIPV1, BIPV2, BIPV3} conversion followed by storage in heat and electricity)\cite{Greenhouse-Heating, SuplimentaryI1, SuplimentaryI2}.

To date, the key technologies applied to greenhouses are focused mainly on transforming diurnal solar energy into heat storage complemented with a smart insulation.  The proposed solutions of solar energy utilization and the reduction of heating power via various combined systems (external thermosolar systems coupled with a storage tank, heating pumps and photovoltaics) \cite{Greenhouse-Heating} show promising solutions but the initial investment  with return of investment are visible in the price of vegetables that can be higher than when using classical electricity heating.

Another key issue until now not taken into account is the solar energy conversion, during  daylight, to internal energy of the air from greenhouse with free convective circulation. Given how it is not appropriate to use forced convection in a greenhouse because the airstreams have a negative effect on the plant growth \cite{SuplimentaryIV}, the only solution remains the free convection for thermal energy transfer.

A second requirement is to maximize the heat transfer from a blackened flat surface under sunlight irradiation to the backside cavity. In this respect, the solution proposed with this device is to increase the backside surface of the solar absorber \cite{radiator} from A to nA by one set of “decorative” elements uniformly distributed along the air stream direction. If in an air-heater with forced convection such elements can induce a transition from laminar to turbulent flow decreasing the efficiency of the thermal energy transfer \cite{Forced-Convection1, Forced-Convection2} in the air-heater with free convection such elements have a minimum effect only in the hydrodynamic resistance \cite{Natural-Convection-Experiment, Natural-Convection-Sim}. The maximum heat transfer in the free convective air heaters is dependent only on the residence time with the hot surface \cite{Roughness-Effect}.

Taking into account the above considerations we propose one cost-effective solution for one solar air-heater to supply  thermal energy by free convection of the air into a greenhouse as well as for other farming application \cite{Solar-Dryer}, as for a supplementary advantage of the air heaters which work upon the principle of natural convection is their independence from external energy sources.

 By comparison with other types of solar air systems glazed \cite{Glazed-Solar1,Glazed-Solar2}, unglazed \cite{Unglazed-Solar1,Unglazed-Solar2}, and with double glazing and double pass air circulation \cite{Single-Double} for a higher yield and relative simple production was chosen the simple glazing design with an effective back cavity. To ensure maximal conversion of solar radiation and efficient heat transfer to air flow the solar absorber was made from blackened aluminum. This decision was justified by the fact that, usually, most energy losses occur through the front cover \cite{Single-Double}, other parts being insulated.

\begin{figure}[p]
 \centering
\includegraphics[height=90mm]{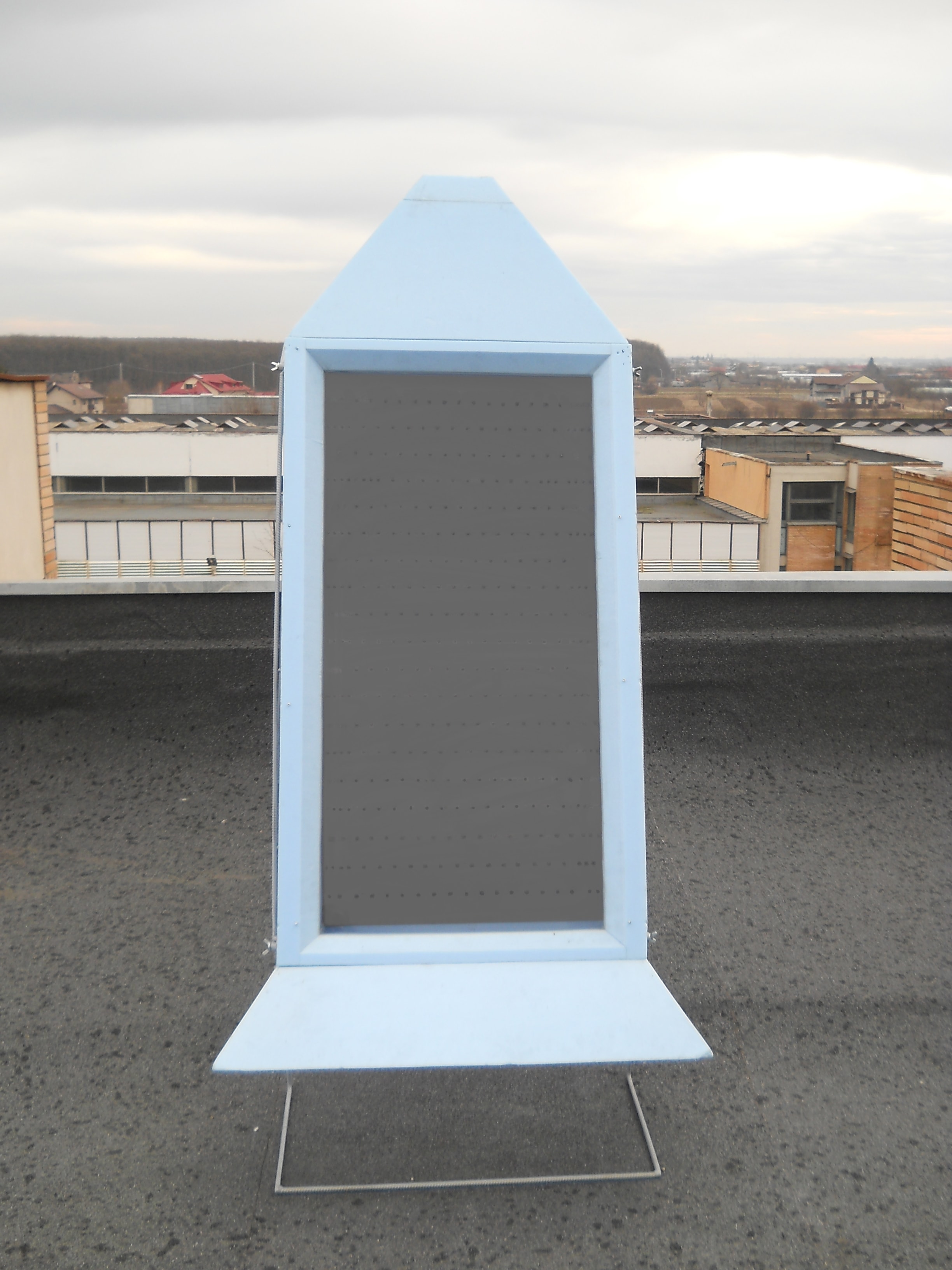}
\caption{Experimental sample of the device.} 
\label{fig:pic0}
\end{figure}

One of the advantages of portable air heaters is the  possibility to flexibly vary the power of the heating in respect to the consumer's preferences and weather. In this paper, we propose a new model of a portable, light-weighted and modular air heater based on solar air heating and natural convection inside of the device shown in Figure \ref{fig:pic0}, which is suitable to be integrated in greenhouses, passive houses and office buildings, and to dry fruits, seeds, and nuts \cite{cashew1, cashew2} during warmer seasons.

\section{Device description. OpenFOAM simulation}

\begin{figure}[p]
 \centering
\includegraphics[height=90mm]{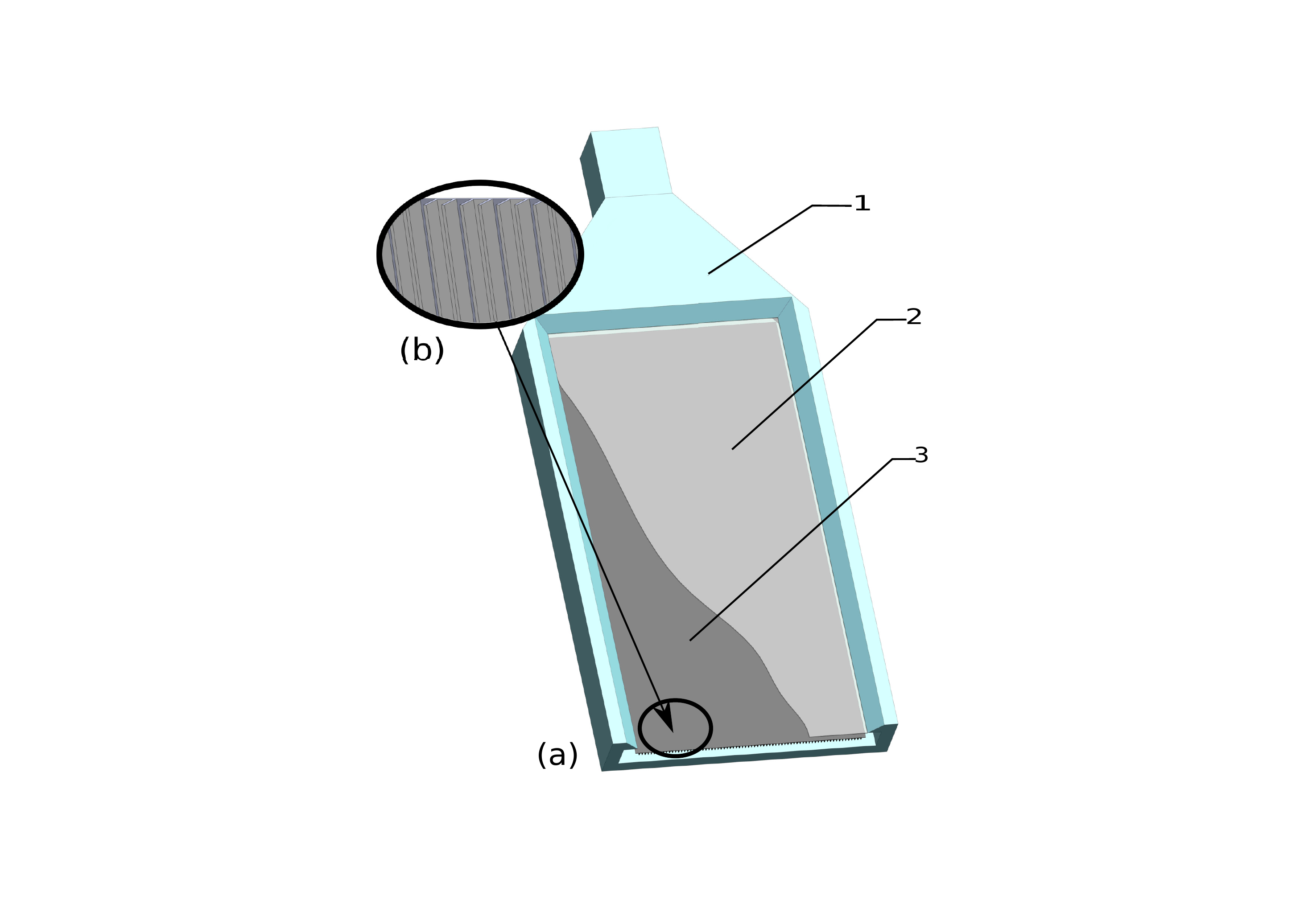}
\caption{Design of the sample of the device. (a) General preview. 1 --- insulated case of the device, 2 --- plexiglass front glazing, 3 --- radiator-like absorbent. (b) Inset shows the detailed U-shape form of profiles of the absorbent part.} 
\label{pic1}
\end{figure}

\paragraph{Solar air heater design} The most effective and flexible air heaters are made as self-contained devices, meaning that they can be attached to an exterior wall of a building \cite{Building-Integration1,Building-Integration2}. A vertical installation is considered. The unit consists of an insulated frame (implemented out of extruded polystyrene foam), a solar absorbent (presented by a blackened aluminum metal board inside of the thermo-insulated case), a front plexiglass glazing, an inlet for the incoming air recirculation, and an outlet. At the outlet of the heater is placed a hood in order to collect hot out-coming air (Figure \ref{pic1} a).

We propose an innovative way of leaving a cavity between the back insulating part and the solar absorbent in order to decrease dissipation of thermal energy. This can lead to an efficiency increase by $30\%$ in comparison to the glazed single pass solar collector, especially in case of air recirculation. In other solar air heaters a part of the thermal energy losses occur due to the contact between hot air and cold front glazing, which we have successfully prevented in our device. For better efficiency, we designed the solar absorbent part to have a flat side that faces the glazing, and the radiator-like surface facing the insulating back part. At the back part of the blackened aluminum board with a thickness of 0.8 mm U-shape profiles are made with the dimensions of 7.5 mm to increase the contact surface  between heated air and the metal (Figure \ref{pic1} b).

Due to the fact that the heat transfer is three times larger at the back side of the metal board in comparison with the front side area \cite{Heat-Transfer-Radiator,radiator}, we have: $Q_{back}/Q_{front}= A_{back}/A_{front} \approx 3 $.

\paragraph{Theoretical model} the functionality of the solar air heater is achieved by the buoyancy phenomena \cite{Buildings-Buoyancy1,Buildings-Buoyancy2} and greenhouse effect \cite{Enegy-Greenhouse}. Solar rays enter the plexiglass glazing and are captured by the solar absorbent. Due to that, the air inside of the installation is heated up. Because of the thermal transfer the air inside of the back cavity is heated up. Temperature differences between the exterior air and the air inside of the heater lead to the appearance of buoyancy force, by which the incoming air is pushed up through the heater.

The decreasing air density because of the heating results in a positive buoyancy force. The greater the thermal difference and the height of the structure, the greater the buoyancy force. At the same time, the maximum value of the velocity is limited by the air friction. The hot air exits the device at the outlet and can be collected for future consumption by a hood.

The airflow used to calculate the efficiency of the air heater can be calculated by the formula: $$\phi=\rho_{out}u_{out}S$$

Where $S$ is the surface area of the hood and $u_{out}$ and $\rho_{out}=\rho_{in} \frac{T_{in}}{T_{out}}$ are velocity and density at the output of the air heater, $T_{out}$ and $T_{in}$ are absolute temperatures at the outlet and inlet of the device respectively.

The thermal performance of the air heater can be calculated by the formula:
\[ \eta = \frac{\phi c_{air} \Delta t}{W_{solar} \cos \theta} \]

where $c_{air}$ is specific heat of air, $\Delta t$ is the difference between inlet and outlet temperatures, $W_{solar}$ is  solar flux measured by pyranometer and $\theta$ is incidence angle.

\paragraph{OpenFOAM numerical simulation} The processes inside the solar converter are strongly nonlinear and cannot be evaluated analytically. For the estimation of the main output parameters we implemented a numerical simulation using OpenFOAM (Open Field Operation and Manipulation). OpenFOAM is an open-source application for computation fluid dynamics (CFD) problems \cite{OpenFOAM-Site,OpenFOAM-Original} with a vast utilization in microclimate research, especially for heat transfer in the air \cite{micro-climat-Open-Foam}.

The solar air heater presented in Figure \ref{pic1} a was approximated by a $2D$ model, that is perfectly justified by the width of the air heater being much greater than its thickness. The ratios between the dimensions of the simulated model correspond to the existing assembled sample of the air heater. Due to the U-shaped profiles at the back part of the solar absorbent, the heat flow is three times higher in comparison with the front side. The external back wall was treated as adiabatic due to the insulation of the air heater. For the transparent window was used constant temperature condition. The radiator is heated up by the sun and correspondingly heats up the air, and the device was studied in the steady state case, therefore the heat flux transfered from the front and back sides of the blackened aluminum board was taken as constant (fixedGradient temperature condition).

\begin{figure}[p]
 \centering
\includegraphics[height=90mm]{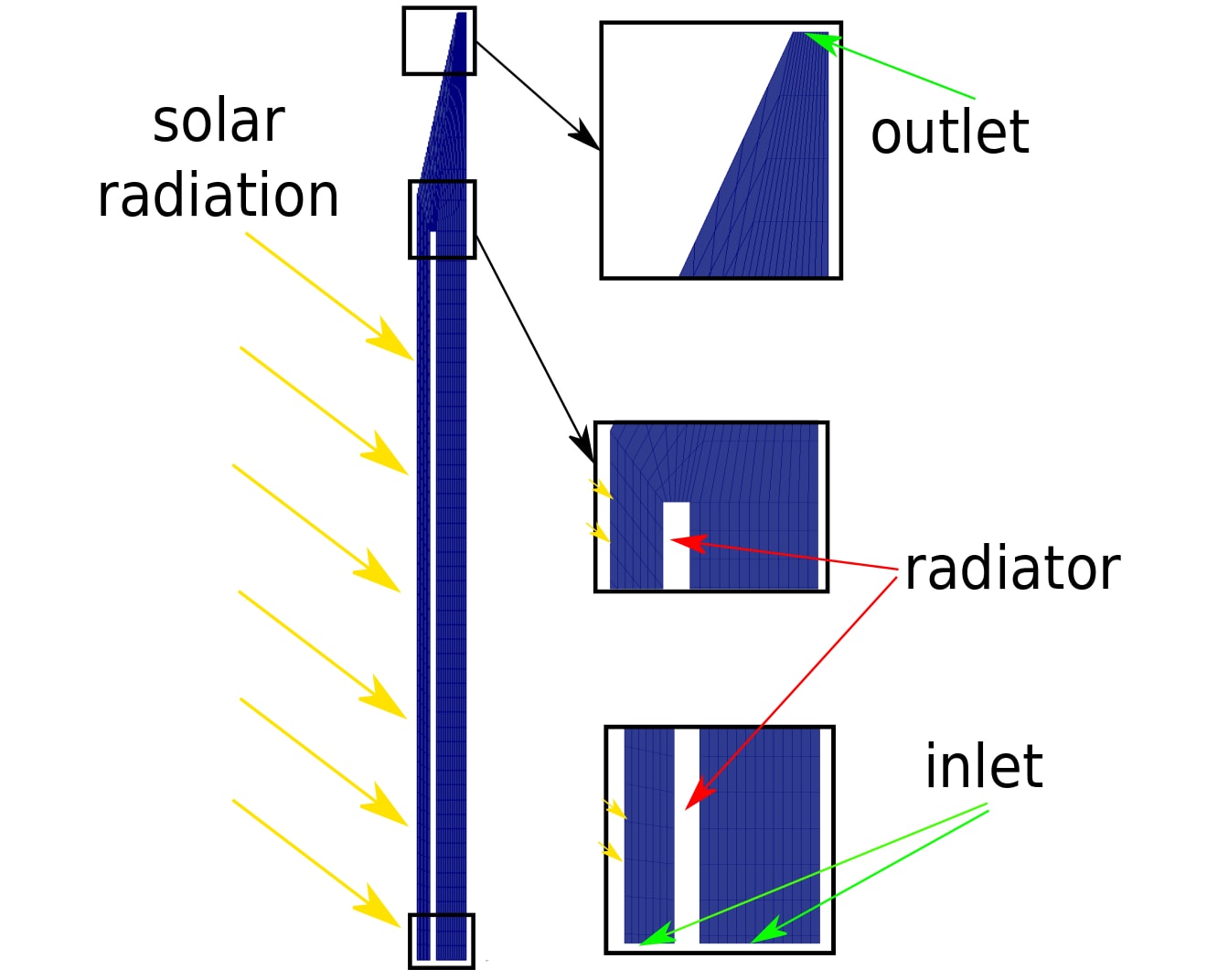}
\caption{Discretization scheme of the air-heater model.} 
\label{pic2}
\end{figure}

The heat produced by the installation depends on the incidence angle ($60^o$ on average) and the absorber efficiency of $90\%$ (due to reflection). The effective total heat flux was considered $600W/m^2$ and the inlet temperature of $27^oC$, which corresponds to the fruit drying process during the warm season. Our tests showed that the inlet temperature is not too relevant for the temperature difference obtained. The discretization was designed by an orthogonal mesh presented in Figure \ref{pic2}. The $y^+$ value was calculated to be between $30$ and $150$ during the whole simulation period. As the initial values for the sample simulations we used the averaged experimental data. For simulation of industrial device, we gradually increased the height of the device and have taken into account the results of previous simulation. Were used k, omega, epsilon and alphat standard wall functions recommended for Buoyant Simple Foam solver.

For the simulation ($k-\omega$) RANS (Reynolds Averaged Nervier–Stokes) turbulence model and Buoyant Simple Foam solver were used, which according to \cite{Open-FOAM-Example} gives satisfactory results. It is a steady state solver, which uses ideal gas approximation for air convection for buoyant and turbulent flow of compressible fluids for ventilation and heat transfer.

The equations used by Buoyant Simple Foam solver are:

\begin{equation*}
 \begin{cases}
\frac{\partial \rho}{\partial t}+\nabla \cdot(\rho u) = 0 
   \\
\frac{\partial}{\partial t}(\rho u)+\nabla \cdot(\rho u)-\nabla [\mu_{eff}(\nabla u+\nabla u^T)-\frac{2}{3}\mu_{eff} \nabla \cdot(u T)]= -\nabla p_d -(\nabla \rho)g
   \\
\frac{\partial}{\partial t}(\rho h)+\nabla \cdot (\rho u h)- \nabla(\alpha_{eff} \nabla h)=u \cdot \nabla p
   \\
\rho = \frac{p}{R T}
   \end{cases}
\end{equation*}

The $h$ is the specific enthalpy; $u$ is velocity, $\alpha$ is air thermal diffusivity, and $\mu$ is dynamic viscosity.

The simulation was stopped when the residuals became constant with values less than $10^{-6}$ for temperature and pressure and $10^{-5}$ for velocity.

The simulation was repeated for different heights of the heater within parameters from $1$ to $20$ meters, and for three different values of solar fluxes, that correspond to a sunny day ($900W/m^2$), an average day  ($600W/m^2$) and a cloudy day ($300W/m^2$).

\section{Results and discussions}

\paragraph{Results of the tested sample} Aiming to obtain the experimental proof regarding the functionality of the device, a testing sample  was built. The thickness of the sample was the same as in the description; the width of $0.56$ m and height of $1$ meter.

During the test period (from $28^{th}$ of September $2015$ to $11^{th}$ of September $2016$) daily experimental data was collected for input and output temperatures, and output air velocity. The solar air heater was protected from wind influence for its optimal functionality. The OMEGA HHF141 rotating vane anemometer was used to measure the output velocity value. The solar radiation intensity was recorded by the Yanishevskii pyrometer.  Two temperature sensors (AS6200) were installed  near the inlet and outlet of the device. All the data was collected automatically every 7.5 minutes throughout the testing period. Temperatures were also periodically measured at the bottom and top parts of hot the radiator place. The radiator temperature was on average $12^oC$ higher than the air temperature in the nearest flow.

\begin{figure}[p]
 \centering
\includegraphics[height=90mm]{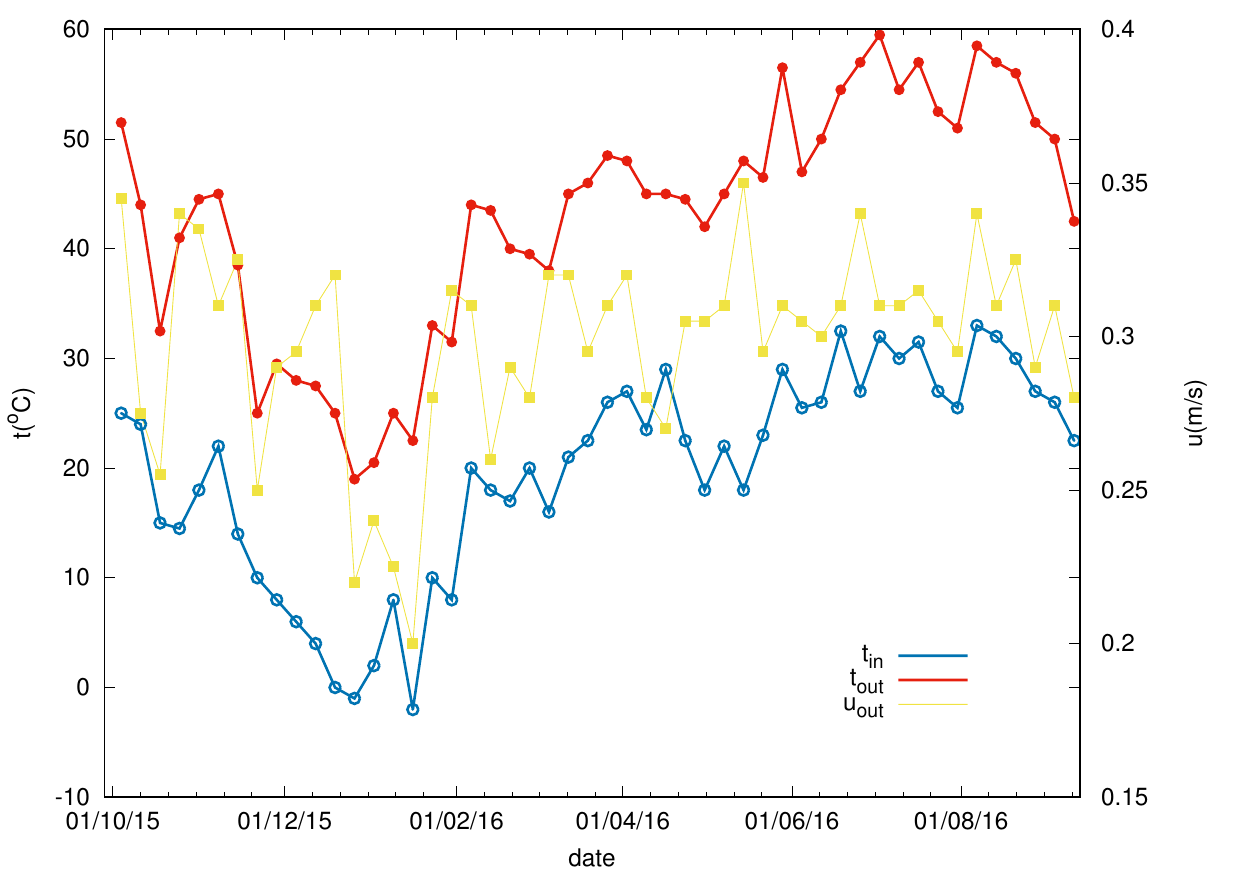}
\caption{Experimental values of input ($t_{in}$) and output ($t_{out}$) temperatures and output velocity ($u_{out}$) .} 
\label{pic3}
\end{figure}

The average experimental data for each week is presented in Figure \ref{pic3}. It was observed that the output temperature depends on the input air temperature and, on average the solar air heater was heating up the air by $23.5^oC$. The mean thermal performance of air heater for the entire testing period was of $\eta =60.4\%$, which was calculated for the mean value of the velocity registered by an anemometer and solar flux being measured by a pyranometer for the sun peak hours. 

\begin{figure}[p]
 \centering
\includegraphics[height=90mm]{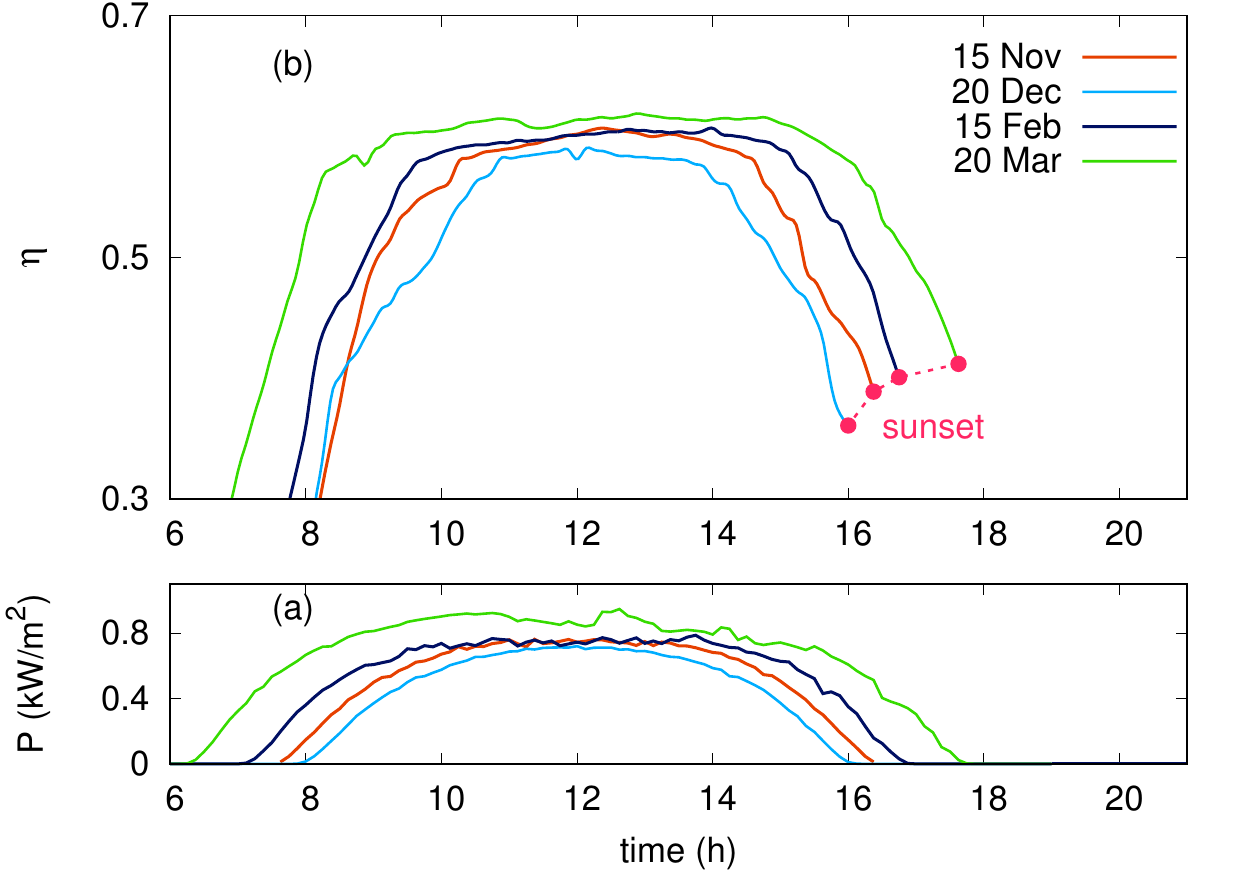}
\caption{(a) The intensity of direct radiation in $kW/m^2$ throughout the day for the dates of the $15^{th}$ of November, the $20^{th}$ of December, the $15^{th}$ of February and the $20^{th}$ of march recorded by the pyranometer. (b) The efficiency of our model of solar air heater throught the corresponding days} 
\label{pic4}
\end{figure}

The variation of the air heater's efficiency (b) and the intensity of the measured solar radiation (a) through the day for the randomly selected colder days of the months is presented in Figure \ref{pic4}. While the $15^{th}$ of November and the $20^{th}$ of December turned out to be some rather clear days, the $15^{th}$ of January and the $20^{th}$ of March were partially cloudy days. Can be observed that the curves for the thermal performance of the testing sample are smoother in comparison to the radiation intensity curves, which is caused by the release of the stored energy by the solar air heater $10$ to $20$ minutes after the receipt. Otherwise, the efficiency of the heater is directly correlated with the received solar energy.
 
The implemented model of the device, as can be seen in Figure \ref{pic4} is more efficient during sunny days. It can be a great supplementary energy source during October, November, February, March, and April. However, it is not efficient enough during the months of December and January. The better effect of the solar air heater can be observed in the second part of the day, when the optimal thermal conditions are the most needed in residence buildings.

\paragraph{Simulation study} CFD numerical simulations offer the detailed image on physical parameters inside of the air heater. This gives us a better understanding of heat transfer and buoyancy that occur in the installation.

\begin{figure}[p]
 \centering
\includegraphics[height=90mm]{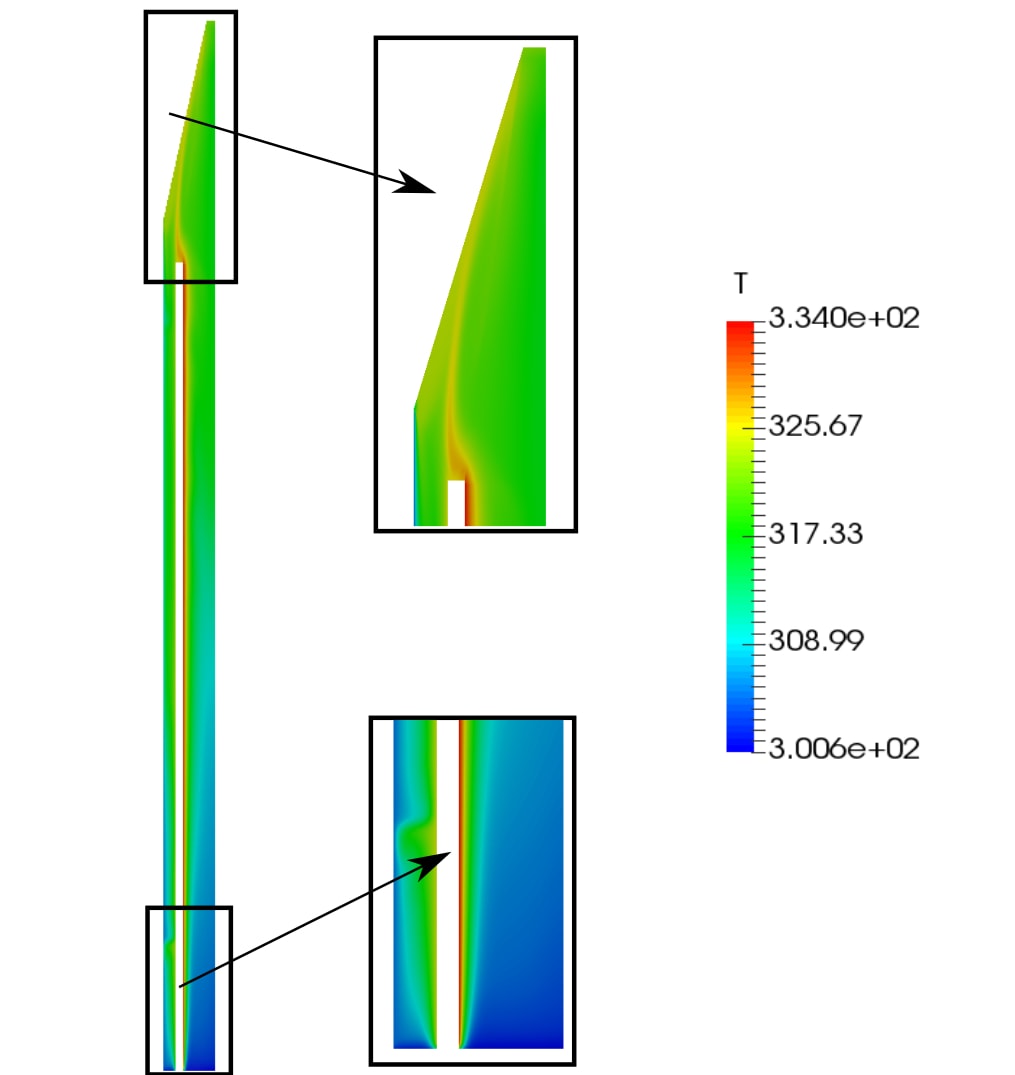}
\caption{Temperature distribution inside of the sample.} 
\label{pic5.a}
\end{figure}

Initially, we built our version of the device in order to establish if there is an agreement of theoretical results with the experimental ones. In Figure \ref{pic5.a} is presented the distribution of the temperature inside the air heater. We can observe that the temperature grows almost linearly along the metal board. A thin layer of air near the metal board appears with a higher temperature in comparison with the average temperature at the same hight. This is due to the fact that the aluminum solar absorbent reaches a temperature $12^oC$ greater than the average air temperature. According to the CFD numerical simulation, the air is heated up by $22^oC$, which is in good agreement with the experimental results, where we obtained heating up by $23.5^oC$.

\begin{figure}[p]
 \centering
\includegraphics[height=90mm]{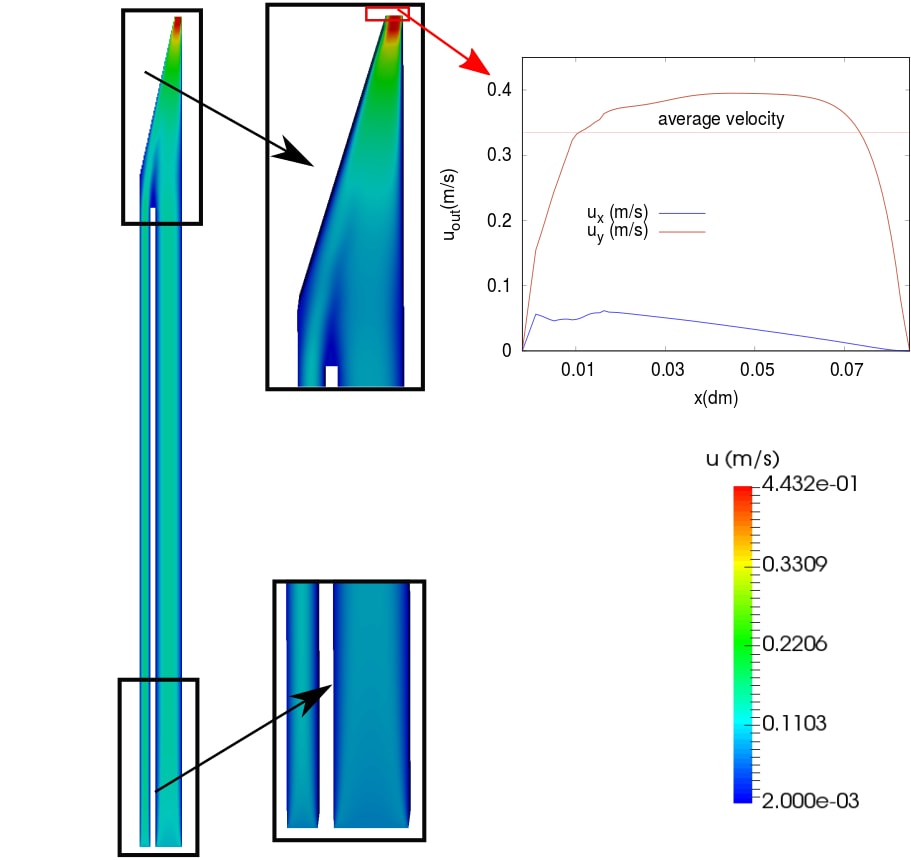}
\caption{Air velocity distribution inside of the sample. The inset with the plot shows the velocity distribution at the outlet along x and y directions and its average value of $0.33$ m/s.} 
\label{pic5.b}
\end{figure}

In Figure \ref{pic5.b} is represented the distribution of velocities inside the air heater. During our work, we obtained good results, the velocity of air being important because it increases the efficiency of the heater, less energy being dissipated in contact with front glazing area. The velocity is almost zero near the metal board and insulated walls, and reaches its maximum value due to the air viscosity between the solar absorbent and the insulated back side. The velocity value increases in the hood area and reaches its maximum at the outlet, due to reduced outlet area in comparison with the inlet one.

Key values of experimental and theoretical results are carried out in the tabular form: \ref{tab1}

\begin{figure}[p]
 \centering
\includegraphics[height=90mm]{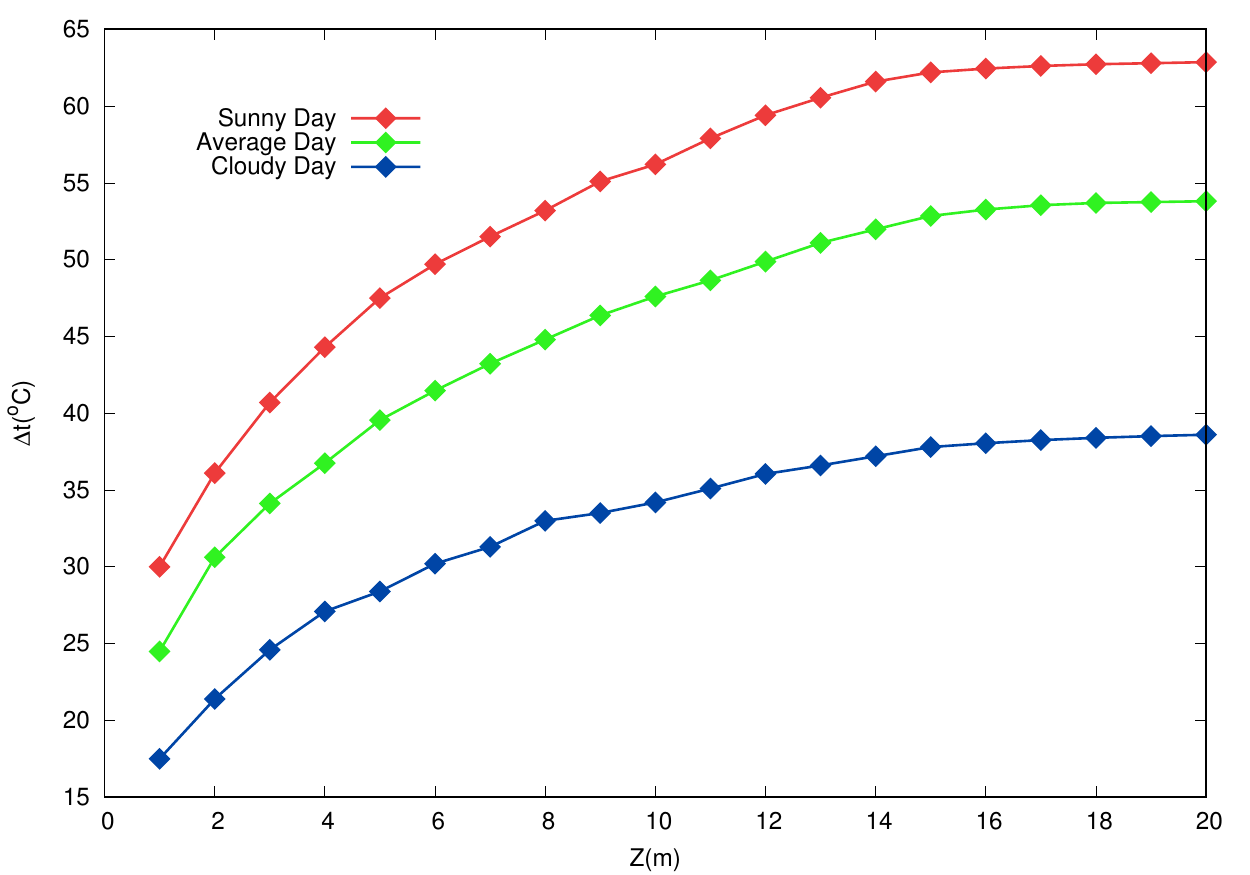}
\caption{Dependence of the difference in temperatures at the inlet and outlet areas on the height of the building for various solar fluxes.} 
\label{pic6.a}
\end{figure}

\begin{figure}[p]
 \centering
\includegraphics[height=90mm]{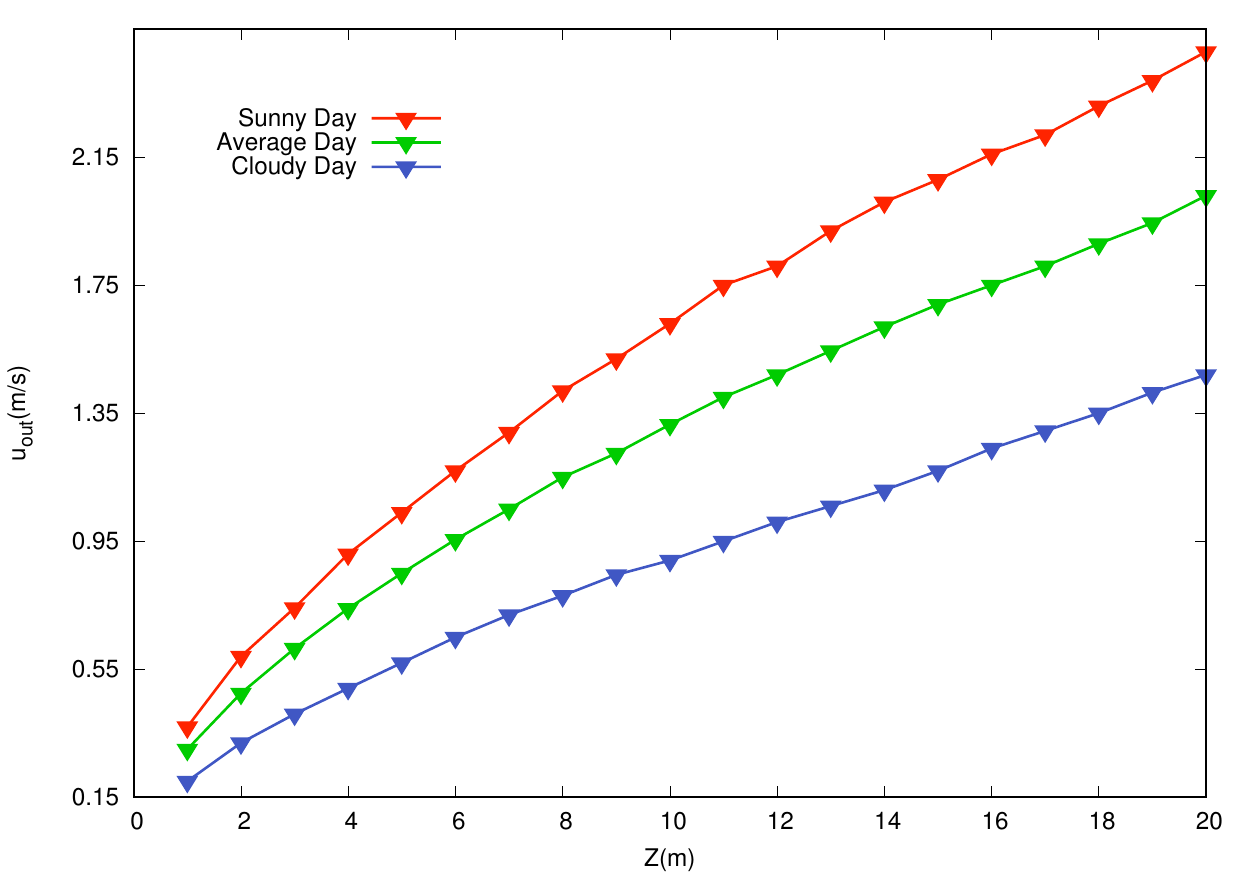}
\caption{Dependence of the output velocity on the height of the building for various solar fluxes.} 
\label{pic6.b}
\end{figure}

After obtaining correspondence of the theoretical and experimental results we decided to run our simulation for different heights of the solar air heater. We propose that the industrial models of the device will have the same thickness, but a larger height and width adopted to building dimensions.  In Figure \ref{pic6.a} is presented the dependence of the output velocity and in Figure \ref{pic6.b} of inlet-outlet temperature difference depending on the height of the heater. We can observe that the output velocity and temperature difference increase with the height of the device. The temperature difference increases slower than the velocity, because thermal dissipation becomes considerable.

The simulation results show that in case of an average day the temperature difference increases from $22^oC$ for $1$ meter to $53.8^oC$ for $20$ meters height of the device. Also we can observe that for heights larger than $12$ meters the temperature increases slower and strives to a limit, which it reaches at about $16$ meters height. That being the optimal height of the device for tall buildings. The velocity increases from $0.33$ m/s for elaborated sample dimensions to $2.03$ m/s for $20$ meters height of the device. For large values of height the velocity increases almost linearly.

\section{Conclusions}

The analysis of numerical simulations and experimental data lead us to conclude that this product can be used as an air heating system for residential and office buildings. It is also very suitable for green-houses heating, due to it working upon the principle of natural convection, to avoid airstreams and their negative effect on plant growth. For thermal comfort of the indoor environment the device can be used as a heating and ventilation system depending on the needs of the user. When outdoor temperatures are high and there is no need for the heating system, the solar air heater can be used to dry fruits, nuts and seeds.

Our solar air heater can replace the heating system during the autumn-spring period when the outdoor temperature is above $-10^oC$. The average thermal performance of the designed air heater is of $\eta=60.4 \%$ for the sun peak hours, and the device can significantly reduce heating costs and protect the environment. The heater can produce a flow of around $59$ l/s of air that will be $23^oC$ hotter then outside temperature. Also, the solar air heater protects walls against temperature fluctuations, provides thermal insulation, and acoustic insulation, which lead to comfort increase.  

\section*{Acknowledgements}

This work was partially supported by the Romanian Ministry of National Education by the contract PN 16 47 0101 with UEFISCDI.

This work was partially supported by the Romanian Ministry of National Education by the contract : PN-II-PT-PCCA-2013-4-1102 with FC-FARM 46-2014

Raluca Gherasim was supported by the strategic grant POSDRU/159/1.5/S/133652, ”Integrated system to improve the quality of doctoral and postdoctoral research in Romania and promotion of the role of science in society” co financed by the European Social Found within the Sectorial Operational Program Human Resources Development 2007 --- 2013.

\section*{References}

\bibliography{mybibfile}

\begin{table}[p]
  \centering
\begin{tabular}{ |l|l|l| }

\hline
 Data & Experimental& Simulation \\ \hline 
$\Delta T$ & $23.5^oC$ & $22^oC$\\ \hline
$\eta$ & $60.4\%$ & $63.9\%$  \\ \hline

 \end{tabular}
  \caption{Key values for air heater}
  \label{tab1}
\end{table}

\end{document}